\documentstyle[12pt]{article}

\def\thefootnote{\fnsymbol{footnote}}
\def\al{\alpha}
\def\be{\beta}

\def\la{\lambda}
\def\sma#1{\mbox{\footnotesize #1}}

\def\Salhbeh{{\cal S_{\widehat{\alpha},\widehat{\beta}}}}
\def\Sab{{\cal{S}}_{_{\!p,q}\!}}
\def\alh{\widehat{\alpha}}
\def\beh{\widehat{\beta}}
\def\11{\mbox{$1$}}

\renewcommand{\thefootnote}{\alph{footnote}}

\newcommand{\rref}[1]{(\ref{#1})}
\newcommand{\beqn}{\begin{equation}}
\newcommand{\eeqn}{\end{equation}}
\newcommand{\beqarr}{\begin{eqnarray}}
\newcommand{\eeqarr}{\end{eqnarray}}

\newcommand{\matc}{\begin{array}{c}}
\newcommand{\matcc}{\begin{array}{cc}}
\newcommand{\matccc}{\begin{array}{ccc}}
\newcommand{\matcccc}{\begin{array}{cccc}}
\newcommand{\emat}{\end{array}}

\newcommand{\Lag}{L}
\newcommand{\Lc}{{\cal L}}
\newcommand{\ul}{\underline}
\newcommand{\IH}{\relax{\rm I\kern-.18em H}}
\newcommand{\IR}{\relax{\rm I\kern-.18em R}}
\newcommand{\IK}{\relax{\rm I\kern-.18em K}}
\newcommand{\II}{\hbox{\rm 1\kern-.35em 1}}
\newcommand{\Is}{\relax{\rm 1\kern-.35em 1}}

\begin{document}

\begin{titlepage}

September 1999         \hfill
\vskip -0.55cm 
\hfill    UCB-PTH-99/37  
\\
\vskip -0.55cm 
\hfill  LBNL-44204    
\begin{center}
%
\renewcommand{\thefootnote}{\fnsymbol{footnote}}
{\large \bf Nonlinear Self-Duality in Even Dimensions}
\vskip .15in
Paolo Aschieri\footnote{email address: aschieri@lbl.gov},
Daniel Brace\footnote{email address: DMBrace@lbl.gov}, 
Bogdan Morariu\footnote{email address: BMorariu@lbl.gov} and 
Bruno Zumino\footnote{email address: zumino@thsrv.lbl.gov}
\vskip .20in

{\em    Department of Physics  \\
        University of California   \\
                                and     \\
        Theoretical Physics Group   \\
        Lawrence Berkeley National Laboratory  \\
        University of California   \\
        Berkeley, California 94720}
\end{center}

\begin{abstract}
We show that the Born-Infeld theory with $n$ complex abelian gauge
fields written in an auxiliary field formulation has a $U(n,n)$ duality
group. We conjecture the form of the Lagrangian obtained by 
eliminating the auxiliary fields
and then introduce a new reality structure leading to a Born-Infeld theory
with $n$ real gauge fields and an $Sp(2n,\IR)$ duality symmetry.
The real and complex constructions are extended to arbitrary
even dimensions.
The maximal noncompact 
\mbox{duality} group is  $U(n,n)$ for complex fields.
For real fields  the duality group is  $Sp(2n,\IR)$ if half of the 
dimension of space-time is even and $O(n,n)$ if it is odd. 
We also discuss duality under the maximal compact subgroup,
which is the self-duality group of the theory obtained by fixing 
the expectation value of a scalar field.
Supersymmetric versions of self-dual theories in four dimensions are 
also discussed.
\end{abstract}

\end{titlepage}

\newpage
\renewcommand{\thepage}{\arabic{page}}

\setcounter{page}{1}
\setcounter{footnote}{0}

\section{Introduction}
\label{intro}
 
Shortly after the appearance of duality in extended supergravity~\cite{FSZ,CJ}
the theory of duality invariance of theories with abelian
gauge fields was developed in~\cite{GZ,BZ}. However, there are very few
examples of duality invariant 
interacting gauge theories where the Lagrangian is 
known in closed form. The most famous is the Born-Infeld 
theory~\cite{BI,Sch,GR1,GR2,GZ1,GZ2} and
in this paper we study its generalization 
to more than one abelian gauge field.

In Section~\ref{theory} we present in some detail the theory
of duality invariance for a theory of complex gauge fields with 
holomorphic duality transformations. 
This is an extension of 
the theory of duality invariance developed in~\cite{GZ,BZ} and
was briefly discussed in~\cite{BMZ}. However, 
the duality group can be larger than that presented in~\cite{BMZ}. In 
fact, for a gauge theory with $n$ complex gauge fields 
the largest possible duality group is $U(n,n)$. We also discuss how to 
obtain such a theory from a theory with a $U(n)\times U(n)$ duality
group, which is the maximal compact subgroup of $U(n,n)$, 
by introducing an additional $n$-dimensional matrix valued scalar field.

In Section~\ref{BIaux} we describe the Born-Infeld Lagrangian
introduced in~\cite{BMZ} and written in terms of auxiliary fields.
Its form is closely related to the Lagrangian introduced 
in~\cite{APT,RT} but differs in two ways. We use a different
reality structure for our fields and introduce a dynamical scalar field 
such that  the duality group is extended to a noncompact group.

In Section~\ref{NoAux} we discuss the elimination of the auxiliary
fields. We have not been able to solve analytically the
nonlinear matrix equations obtained from the variation of the auxiliary
fields. However we have calculated the first few terms in the 
perturbative expansion of the Lagrangian in the field strength and
based on these we have conjectured in~\cite{BMZ} the form of the
Lagrangian to all orders. 
In~\cite{BMZ} the conjecture was checked by hand 
up to the sixth order. It has now been checked 
by computer up to the seventeenth order. In Appendix A
we discuss an equivalent perturbative expansion of the Lagrangian
which simplifies the order by order check  of the
conjecture.
  
In the theory with auxiliary fields it does not seem possible to 
work with real gauge fields, but this can  be done 
in the Lagrangian with the  auxiliary
fields eliminated. As will be shown in Section~\ref{RealF} this leads to a
Born-Infeld theory with an $Sp(2n,\IR)$ duality group. 
Assuming that the conjecture of Section~\ref{NoAux} is correct,
this would be the first example of an interacting gauge theory
whose Lagrangian is known to all orders and whose 
duality group is as large as the duality group of
the Maxwell theory with the same number of gauge fields. 

In Section~\ref{SUSY}  we show how to supersymmetrize the
Born-Infeld Lagrangian in the formulation with auxiliary fields. 
We also present  the form without auxiliary fields of
the supersymmetric Born-Infeld Lagrangian with a single gauge field
and a scalar field; this theory is invariant under $SL(2, \IR)$
duality, which reduces to $U(1)$ duality if the value of the scalar
field is suitably fixed. Versions of this theory without the scalar
field were presented in~\cite{DP,CF,BG}.

In Section~\ref{forms} we generalize our construction to arbitrary even 
dimensions by using antisymmetric tensor 
fields such that the rank of their field 
strength equals half the dimension of space-time. We consider first 
theories with a $U(n,n)$ duality group using complex
antisymmetric tensor fields; then we
discuss theories with real antisymmetric tensor fields. These have an 
$Sp(2n,\IR)$ duality group 
if half of the space-time dimension is even 
and $O(n,n)$ if it is odd. The fact that the duality group depends 
on half the dimension of space-time was discussed earlier 
in~\cite{tanii,Ferrara,Julia,Julia2,araki}.

Finally in Appendix B we briefly discuss two parametrizations of the
coset space $U(n,n)/U(n)\times U(n)$  and show how the left
multiplication on  $U(n,n)$ induces fractional transformations in one
of the parametrizations. We also discuss the corresponding coset
spaces of the symplectic and orthogonal group and describe their
global structure.

\section{Duality Invariance}
\label{theory}

In this section we describe how the theory of self-duality
introduced in~\cite{GZ, BZ} is modified when we 
consider complex abelian gauge fields. 
We only consider
a linear action of the duality group which mixes the field strengths
and their duals but not their complex conjugates. We will refer to
this as a holomorphic action.  
Under these conditions the largest allowed duality group is $U(n,n)$
where $n$ is the number of complex gauge fields.  
If we do not require a holomorphic action, $n$ complex gauge fields are 
equivalent to $2n$ real gauge
fields in which case the  
largest possible duality group is $Sp(4n,\,\IR)$. 
Later, in Section~\ref{RealF}, we will also 
introduce a Born-Infeld action with real
gauge fields which we conjecture to have the largest allowed duality group 
given the number of gauge fields. However, the 
argument leading to this conjecture involves Lagrangians with complex
gauge fields.  

Consider a theory of  $n$ complex abelian gauge fields 
and a  scalar field $S$ which is an $n$-dimensional complex matrix. 
Here we do not require $S$ to be symmetric and
as a result we find a larger duality group 
than the one appearing in~\cite{BMZ}.  
The gauge fields only enter in the Lagrangian 
through their field strengths $F^a$, where $a=1,\ldots,n$, 
and their complex conjugates~$\bar{F}^a$
\beqn
\Lag= \Lag(F^a,\bar{F}^a,S,\ldots)~.\label{LLL}
\eeqn
The dots in~\rref{LLL} represent possible 
auxiliary fields which could also be present in $\Lag$.
As we will show later, with the scalar field $S$ 
present the duality group is noncompact while 
without the scalar field only the maximal compact subgroup survives.
We can also add to this Lagrangian a kinetic term for the scalar field
$S$.
As explained in~\cite{GZ} additional physical fields, e.g. spinors, 
can also be introduced, but we shall not consider them explicitly in
this paper except in Section~\ref{SUSY} where the supersymmetric
Born-Infeld theory is discussed.
 
The dual field strength, or rather the Hodge dual of the dual field strength,
$\widetilde{G}_{\mu \nu}^a=
\frac{1}{2}\varepsilon_{\mu\nu\rho\sigma}G^{a\,\rho\sigma}$,  is defined as
\beqn
\tilde{G}^{a}_{\mu\nu}\equiv 2\frac{\partial \Lag}
{\partial  \bar{F}^{a\,\mu\nu}}~,~~
\tilde{\bar{G}}^{a}_{\mu\nu}\equiv 2\frac{\partial \Lag}
{\partial  F^{a\,\mu\nu}}~.
\label{Gdef}
\eeqn
Throughout this paper we will assume that we are in four space-time 
dimensions, except in Section~\ref{forms}, where  we will show how to
generalize our results to theories in even space-time dimensions.  

The equations of motion and Bianchi identities transform  covariantly 
under the 
following holomorphic infinitesimal transformations
\beqn
\delta
\left(
\matc
G \\
F
\emat
\right)
=
\left(
\matcc
A & B \\
C & D
\emat
\right)
\left(
\matc
G \\
F
\emat
\right)~.\label{FGtrans}
\eeqn
Let $\phi$\, denote all the scalar fields appearing in the Lagrangian and
$\phi_{\mu}=\partial_{\mu} \phi$\,. The infinitesimal transformations
of the scalar fields are given by
\beqn
\delta \phi^i =\xi^i (\phi)~,
\label{Phitrans}
\eeqn
where $\xi^i$ are components of a vector field on the scalar field space.
The most general Lagrangian, neglecting possible fermionic fields, has
the form $L(F,\,\bar{F},\,\phi,\,\phi_{\mu})$\,. Its variation 
under~\rref{FGtrans}\rref{Phitrans} can be written as
\[
\delta L=
\left[
\delta_{\phi} +
(G C^T +F D^T)\frac{\partial}{\partial F} +
(\bar{G} C^{\dagger}+\bar{F}D^{\dagger})\frac{\partial}{\partial \bar{F}}
\right] \,L~,
\]
where $\delta_{\phi}\, L$ is given by
\[
\delta_{\phi}\, L=
(
\xi^i\, \frac{\partial}{\partial \phi^i}
+
\phi^j_{\mu}\,
\frac{\partial \xi^i}{\partial \phi^j}\,
\frac{\partial}{\partial \phi^i_{\mu}}
)\, L~.
\]
The variation of the Lagrangian must satisfy certain consistency
conditions. First note that 
\[
\frac{\partial}{\partial F}\,\left( \delta L\right)=
\delta \left(\frac{\partial L}{\partial F}\right) +
\frac{\partial G}{\partial F}\,
C^T\,
\frac{\partial L}{\partial F}
+
D^T\, \frac{\partial L}{\partial F}+
\frac{\partial\bar{G} }{\partial F} \,C^{\dagger}\,
\frac{\partial L}{\partial \bar{F}}~.
\]
Using~\rref{Gdef} we obtain 
\beqn
\delta\, \widetilde{\bar{G}}=
2\frac{\partial}{\partial F} (\delta  L)
-\bar{G} \,C\, \frac{\partial \widetilde{G}}{\partial F} -
\frac{\partial \bar{G}}{\partial F} \,C^{\dagger} \,\widetilde{G}
-\widetilde{\bar{G}}\, D~,
\label{dG1}
\eeqn
and this should be consistent with the variation obtained from~\rref{FGtrans}
\beqn
\delta\, \widetilde{\bar{G}}
= \widetilde{\bar{G}}\,A^{\dagger} +\widetilde{\bar{F}}\,B^{\dagger}~.
\label{dG2}
\eeqn
Equating~\rref{dG1} and~\rref{dG2}  we obtain the consistency condition
\beqarr
\frac{\partial}{\partial F}\,
\left( \delta L      
-\frac{1}{4} \bar{G} (C^{\dagger} + C)\widetilde{G}
-\frac{1}{4} \bar{F} (B^{\dagger} + B)\widetilde{F}
 \right)=~~~~~~~~~~~~~~~
\label{CC1}
\\
\frac{\partial L}{\partial F } (D+A^{\dagger})
+
\frac{1}{4}
\bar{G}(C-C^{\dagger})
\frac{\partial \widetilde{G}}{\partial F}
-\frac{1}{4}
\frac{\partial \bar{G}}{\partial F}
(C-C^{\dagger}) G+
\frac{1}{4}\widetilde{\bar{F}}(B^{\dagger} -B)~.
\nonumber
\eeqarr
The right hand side of the above equation must be a total derivative
since the left hand side is one. This is possible if 
\[
A^{\dagger}+D=\varepsilon I~,~~B^{\dagger}=B~,~~C^{\dagger}=C~,
\]
where $\varepsilon$ is a real parameter. 
These are the relations of the fundamental representation of 
the $U(n,n)\times \IR^{\star}$\, Lie 
algebra\footnote{$\IR^{\star}$ denotes the group of nonvanishing real numbers.}.
We will only consider the
case when $\varepsilon$ vanishes. Thus we assume
\beqn
A^{\dagger}= -D~,~~B^{\dagger}=B~,~~C^{\dagger}=C~.
\label{ABrel}
\eeqn
The relations~\rref{ABrel} define
the fundamental representation of the Lie algebra of
$U(n,n)$. However, in general the transformations~\rref{Phitrans} 
of the scalar fields can be implemented only for a subgroup $H$ of
$U(n,n)$. The duality group $H$ depends both on the field content and the
nature of the interactions of the scalar fields.

Using~\rref{ABrel} the consistency condition~\rref{CC1} can be written as
\beqn
\frac{\partial}{\partial F}
\left(
\delta L -\frac{1}{2} \bar{F} B \widetilde{F} 
 -\frac{1}{2} \bar{G} C \widetilde{G} 
\right)
=0~.
\label{Consist1}
\eeqn
Another consistency condition is obtained by applying the
Euler operator
\[
 \frac{\partial}{\partial \phi^i} - \partial_{\mu}
\frac{\partial}{\partial \phi^i_{\mu}} 
\]
on the variation of the Lagrangian. Similarly to a derivation in~\cite{GZ},
and assuming~\rref{ABrel} we obtain
\beqn
 \left(\frac{\partial}{\partial \phi^i} - \partial_{\mu}
\frac{\partial}{\partial \phi^i_{\mu}}
\right)
\left( \delta L- \frac{1}{2} \bar{F} B \widetilde{F} 
 -\frac{1}{2} \bar{G} C \widetilde{G} \right)=
\delta E_i +\frac{\partial \xi^j}{\partial \phi^i}\, E_j~,
\label{Consist2}
\eeqn
where $E_i$ is the left hand side of the 
equation of motion for the field
$\phi^i$\,
\[
E_i=\frac{\partial L}{\partial \phi^i} - \partial_{\mu}
\frac{\partial L}{\partial \phi^i_{\mu}}~.
\]
A sufficient condition to satisfy the consistency
equation~\rref{Consist1} is given by
\beqn
\delta \Lag= \frac{1}{2}(\bar{F} B\widetilde{F} +\bar{G}
C\widetilde{G})~.
\label{dLag}
\eeqn
This is equivalent to the invariance of the following combination
\beqn
\Lag-\frac{1}{4}\bar{F} \widetilde{G}-\frac{1}{4} F
\widetilde{\bar{G}}~.
\label{inv}
\eeqn
Using~\rref{dLag} in~\rref{Consist2} we obtain 
\beqn
\delta E_i =-\frac{\partial \xi^j}{\partial \phi^i}\, E_j~,
\label{CovScalars}
\eeqn
showing that the equations of motion for the scalar fields 
form a multiplet under the duality group $H$. In the examples
discussed in this paper the duality group will be $U(n,n)$ for complex 
gauge fields and $Sp(2n, \IR)$ for real gauge fields. Ignoring a
possible $\IR^{\star}$ factor, present only for a nonvanishing $\varepsilon$, 
we will refer to these as the maximal noncompact duality groups.

The corresponding finite duality transformations are given by
\beqn
\left(
\matc
G' \\
F'
\emat
\right)=
M
\left(
\matc
G \\
F
\emat
\right)~.
\label{Fintrans}
\eeqn
Here $M$ is an $U(n,n)$ matrix satisfying 
\beqn
M^{\dagger}\, \IK \,M = \IK~,
\label{MKMK}
\eeqn
where $M$ and $\IK$ have the block form
\[
M=
\left(\begin{array}{rr}
a & b\\
c & d
\emat
\right)~,~~
\IK
=
\left(\begin{array}{rr}
0 & 1\\
-1 & 0
\emat
\right)~.
\]
Note that the invariant $\IK$ defining $U(n,n)$ is the usual off
diagonal symplectic form. This explains the similarity of our results
with the real case discussed in~\cite{GZ}.
One can check that~\rref{MKMK} implies the following  relations for 
the block components of $M$
\beqn
c^{\dagger} a = a^{\dagger} c~,~~ b^{\dagger} d = d^{\dagger} b~,
~~ d^{\dagger} a-b^{\dagger} c =1~.
\label{sp2n}
\eeqn
The infinitesimal relations~\rref{ABrel} can be obtained from the
finite relations~\rref{sp2n} using
\[
a\approx 1+A~,~~b \approx B~,~~c\approx C~,~~d\approx 1+D~.
\]

In much of this paper we consider Lagrangians which do not depend on
the scalar field $S$, i.e. they depend only on 
the gauge field strengths and perhaps auxiliary scalar fields, 
and are invariant only under the 
maximal compact subgroup $U(n)\times U(n)$ of 
$U(n,n)$. 
Then there is a way to introduce the scalar field $S$ 
which extends the duality group to $U(n,n)$.
The maximal compact subgroup $U(n)\times U(n)$
is the subgroup of $U(n,n)$  obtained by requiring~\rref{sp2n} and
\[ 
a= d~,~~ b= -c~.
\]
The corresponding infinitesimal relations are~\rref{ABrel} and
\[ 
A= D~,~~ B= -C~.
\]
Let $\Lc(F,\bar{F})$ be a Lagrangian describing a theory invariant
under  $U(n)\times U(n)$, where we suppress the dependence on the 
auxiliary fields.
Then we define a new Lagrangian
\beqn
\Lag (F,\bar{F},S_1,R,R^{\dagger}) \equiv
\Lc (RF,\bar{F}R^{\dagger}) + \frac{1}{2}\,{\rm Tr}(S_1  \tilde{F}\bar{F})~,
\label{extension}
\eeqn
where  $S_1$ is a hermitian $n$-dimensional matrix and $R$ is a
nondegenerate  $n$-dimensional matrix. This Lagrangian
describes a theory invariant under $U(n,n)$ if we transform the scalar 
fields $S_1$ and $R$ as discussed below. 
As we will see, the duality invariance of the
theory described by $\Lc$ 
implies that $\Lag$ depends on $R$ and $R^{\dagger}$  
only through the hermitian positive definite matrix
\beqn 
S_2 = R^{\dagger}R~.
\label{S2rr}
\eeqn 
We also define 
$S \equiv S_1 + iS_2$.
Under the duality group  $S$ transforms by fractional transformation
\beqn
S' = (aS+b)(cS + d)^{-1}~,
\label{dualSS}
\eeqn
whose infinitesimal form is 
\beqn
\delta S= B+ AS-SD-SCS~.
\label{duality99}
\eeqn
It is also convenient to write down the transformation of $S_2$
\beqn
S_2'  =(cS+d)^{-\dagger} S_2(cS+d)^{-1}~.
\label{S2trans}
\eeqn
In~\rref{S2trans} and below we use the notation
$-\dagger$\, for the  hermitian conjugate of the inverse.
 
Next we show that the Lagrangian $L$ defined in~\rref{extension} 
corresponds to a $U(n,n)$ duality invariant theory. We
follow closely~\cite{araki} where the case of 
real gauge fields was considered. The proof in~\cite{araki} generalizes 
the introduction of a single complex scalar field for a $U(1)$
interacting theory discussed in~\cite{GZ1,GZ2}.
Using the fact that $\Lc(F,\bar{F})$ satisfies~\rref{dLag} with compact duality
rotations we have
\beqn
\bar{G}^a \widetilde{G}^b + \bar{F}^a \widetilde{F}^b = 0~,
\label{GG}
\eeqn
\beqn
\bar{G}^a \widetilde{F}^b -  \bar{F}^a\widetilde{G}^b 
=0~.
\label{GF}
\eeqn
The relation~\rref{GG} corresponds to transformations with $A=0$ 
while~\rref{GF}
is obtained by setting $C=0$. 
We now introduce some convenient notation
\beqn
{\cal F} = RF~,~~\widetilde{ \bar{{\cal G}}} = 
2 \frac{\partial \Lc ({\cal F},\bar{\cal F})}{\partial{\cal F}}~.
\label{FRF}
\eeqn
Given a Lagrangian $\Lc$ which depends on $F$ but not its 
derivatives, 
we may rewrite~\rref{GG} and~\rref{GF} as
\beqn
\bar{\cal G}^a \widetilde{\cal G}^b + \bar{\cal F}^a \widetilde{\cal F}^b = 0~,
\label{gg}
\eeqn
\beqn
\bar{\cal G}^a \widetilde{\cal F}^b -  \bar{\cal F}^a\widetilde{\cal G}^b 
=0~.
\label{gf}
\eeqn
We would like to show that under an infinitesimal $U(n,n)$ duality
transformation the change in the Lagrangian  $\Lag$ defined 
in~\rref{extension} satisfies the duality condition~\rref{dLag}
\beqn
( \delta_F +\delta_{\bar{F}} + \delta_{S_1} +\delta_R + \delta_{R^{\dagger}})
\Lag = \frac{1}{2}( \bar{F}B\widetilde{F} + \bar{G}C\widetilde{G} )~.
\label{vars}
\eeqn

A transformation law for $R$ which is consistent with the relation 
$R^{\dagger}R = S_2$ and the duality transformation~\rref{S2trans} of $S_2$ 
is given by
\[
R' = R(cS + d)^{-1},
\]
whose infinitesimal transformation is $\delta R = - R(CS + D)$\,.
This choice is somewhat arbitrary since equation~\rref{GF}
is equivalent to the Lagrangian ${\cal L}$ being invariant
under left multiplication of the gauge field strength by unitary
matrices $U$
\[
\Lc (U F,\bar{F}U^{\dagger})=\Lc (F,\bar{F})~.
\]
This ensures that left multiplication of $R$ by a unitary matrix leaves the
Lagrangian $\Lag$ invariant.
It follows that the Lagrangian $\Lag$ 
only depends on $S_2$ and not on the specific
$R$ chosen\footnote{Note that $S_2$ is a positive definite hermitian 
metric and $R$ is a
vielbein. The Lagrangian only depends on the metric and the
arbitrariness in the choice 
of vielbein introduces a $U(n)$ gauge invariance.}, 
as we have already mentioned.
Any variation of the form $\delta R = \Omega R - R(CS+D)$, where $\Omega$ is 
anti-hermitian, would still preserve the relation $R^{\dagger}R =
S_2$.

Using the above transformation of $R$ one can show that~\rref{vars} is
equivalent to the vanishing of the following expression
\[
\widetilde{G}^a \bar{G}^b -\widetilde{G}^a (\bar{F}S_1)^b - 
(S_1 \widetilde{F})^a \bar{G}^b + (S_1 \widetilde{F})^a(\bar{F}S_1)^b +
(S_2 \widetilde{F})^a(\bar{F} S_2)^b +
\]
\[
-i\left((S_2 \widetilde{F})^a \bar{G}^b - (S_2 \widetilde{F})^a (\bar{F}S_1)^b
- \widetilde{G}^a(\bar{F} S_2)^b + (S_1 \widetilde{F})^a (\bar{F} S_2)^b
\right).
\]
Using the relation ${\cal G} = R^{-\dagger}(G - S_1 F)$, which follows
from~\rref{Gdef} and~\rref{FRF},
the first and second lines of this expression are equivalent to
the left hand side of~\rref{gg}
and~\rref{gf} respectively.
Thus~\rref{vars} is satisfied concluding the proof that 
the theory with the Lagrangian $\Lag$ 
is invariant under $U(n,n)$. 

Conversely, if we are given a Lagrangian $\Lag$ with equations of motion 
invariant under  $U(n,n)$ we can obtain a theory without the 
scalar field $S$ by 
setting $S=i$. Then the duality group is broken 
to the stability group of $S=i$ which is $U(n)\times U(n)$, 
the maximal compact subgroup. 
Thus we can easily move between the theory with a scalar field $S$ and
the theory without~$S$. 

We also give the infinitesimal transformation of ${\cal F}$ and ${\cal G}$ 
\beqarr
\delta {\cal G} 
&=& ~~~RCR^{\dagger} {\cal G} -i RCR^{\dagger} {\cal F}~, \label{dFG} \\
\delta {\cal F} 
&=& -RCR^{\dagger} {\cal F} -i RCR^{\dagger} {\cal G}~. \nonumber
\eeqarr
The last term in~\rref{dFG} is a unitary transformation
and could be canceled by  using a different choice for the transformation 
of $R$. The first term is an infinitesimal duality transformation
belonging to the maximal compact subgroup $U(n)\times U(n)$. Note
however that it is a space-time dependent duality transformation. 

Next we find the differential equation that a Lagrangian must
satisfy if the equations of motion are invariant under the maximal 
compact duality group. 
We are therefore considering a Lagrangian without the scalar field~$S$.
We will also assume that the auxiliary fields have been
eliminated, the field strengths appear in the Lagrangian
only through the Lorentz invariant combinations
\beqn
\alpha^{ab} \equiv \frac{1}{2} F^{a} \bar{F}^{b},
~~\beta^{ab} \equiv 
\frac{1}{2}  \widetilde{F}^{a} \bar{F}^{b},
\label{alphadef}
\eeqn
and that the Lagrangian is a sum of traces (or of products of traces)
of monomials
in $\alpha$ and $\beta$.  
If the  Lagrangian has such a form, equation~\rref{GF} is
satisfied.
Then under a compact duality transformation the variation of the Lagrangian is
\[
\delta \Lc = {\rm Tr} ( \Lc_{\alpha} \delta \alpha + 
\Lc_{\beta} \delta \beta)~,
\]
where we define
\[ 
\Lc_{\alpha} \equiv \frac{\partial \Lc}{\partial \alpha^T}~,~~
\Lc_{\beta} \equiv \frac{\partial \Lc}{\partial \beta^T}~.
\]
Using the definitions~\rref{Gdef} and~\rref{alphadef},
we find that~\rref{GG} is equivalent to
\beqn
\Lc_{\beta} \beta \Lc_{\beta} -\Lc_{\alpha} \beta \Lc_{\alpha} +
\Lc_{\alpha} \alpha \Lc_{\beta} + \Lc_{\beta} \alpha \Lc_{\alpha} +
\beta = 0~.\label{bselfdual}
\eeqn
This is a generalization of the differential equation introduced 
in~\cite{GZ2} where the case of a single real gauge field
was considered. Equation~\rref{bselfdual} is invariant under
the following transformation
\beqarr
\alpha'&=&\alpha~, \\
\beta' &=&-\beta~.  \nonumber
\eeqarr
 
If one considers a  self-dual theory with $n$ real field strengths $F_R$, 
where now $\alpha$ and $\beta$ are defined by 
$\alpha^{ab}= 1/4~ F_R^a F_R^b$ and $\beta^{ab}= 1/4~ F_R^a \tilde{F_R^b}$,
equation~\rref{bselfdual} still  holds.
In this case one can
extend the duality group from $U(n)$ to $Sp(2n,\IR)$
by introducing scalar
fields as in~\cite{araki}. 
Although these remarks will be central in later arguments,
their proofs closely resemble those in the case of complex fields,  
so we omit them.

\section{Born-Infeld with Auxiliary Fields}
\label{BIaux}
In this section we describe a $U(n,n)$ duality invariant 
nonlinear gauge theory with $n$ complex gauge fields~\cite{BMZ}.
The use of auxiliary fields in the Lagrangian is inspired by the 
work of~\cite{APT,RT} and simplifies the check of duality invariance. 

We begin with the following Lagrangian introduced in~\cite{BMZ}
\beqn
\Lag={\rm Re}\,{\rm Tr}\,[\,
i(\lambda-S)\chi -\frac{i}{2} \lambda \chi S_2 \chi^{\dagger}
+i\lambda \,{\cal N}\,]~,\label{BIL}
\eeqn
where ${\cal N}=\alpha - i \beta$\,.
As mentioned in Section~\ref{theory}, here we do not require $S$
to be symmetric. 
The auxiliary fields $\chi$ and $\lambda$ are $n$ 
dimensional complex matrices.  
If we could solve their equations of motion and use the solution in
the Lagrangian~\rref{BIL} we would
find a Lagrangian which depends only on $\alpha$, $\beta$ and $S$.
Obtaining this Lagrangian is the main thrust of our paper.

If we set $S=i$ in the above Lagrangian, the theory is only self-dual 
under the maximal compact subgroup $U(n) \times U(n)$, as discussed 
in Section~\ref{theory}. However, if we now reintroduce the scalar field
as in~\rref{extension}, the new Lagrangian is the same 
as~\rref{BIL} only after field redefinitions of $\chi$ and
$\lambda$.
We can also add a kinetic term for the scalar field $S$. This term
must be duality invariant since, as we will see shortly, 
the rest of the Lagrangian already 
satisfies the self-duality condition~\rref{dLag}. 
For example we can add a nonlinear $\sigma$-model 
Lagrangian defined 
on the coset space $U(n,n)/U(n)\times U(n)$ with the metric given by
\beqn
{\rm Tr}\left[
(S^{\dagger}-S)^{-1}d S^{\dagger}(S-S^{\dagger})^{-1}d S
\right].
\label{metric}
\eeqn
The metric~\rref{metric} is K${\rm \ddot{a}}$hler since it is obtained 
from the
K${\rm \ddot{a}}$hler potential 
\beqn
K={\rm Tr\, ln}(S_2)~.
\label{kh}
\eeqn
This K${\rm \ddot{a}}$hler potential changes by a  
K${\rm  \ddot{a}}$hler
transformation under~\rref{duality99};
this ensures that the metric is duality
invariant.

It will be convenient to decompose the auxiliary fields into 
hermitian matrices, as we have already done for $S$,
\[  
S=S_1 + i S_2~,~~\lambda=\lambda_1+i \lambda_2~,~~\chi=\chi_1+i \chi_2~.
\]

To prove the duality of (\ref{BIL}) we first note 
that the last term in the Lagrangian can be written as
\[
{\rm Re}\,{\rm Tr}\,[\,i\lambda (\alpha - i \beta)\,]=
{\rm Tr}(-\lambda_2 \alpha + \lambda_1\beta)~.
\]
If the field $\la$ transforms by fractional transformation and 
the $\la_i$'s and the gauge fields are real this is the $U(1)^n$
Maxwell action, with the gauge fields interacting with the scalar
field $\la$, and this term by itself has the correct
transformation properties under the duality group~\cite{GZ}. 
Similarly for hermitian $\alpha$, $\beta$ and $\lambda_i$ this term by
itself satisfies equation~\rref{dLag}.
It follows that the rest of the Lagrangian must be duality invariant.
The duality transformations of the scalar and auxiliary
fields are
\beqarr
S'         &=&(aS+b)(cS+d)^{-1}, \label{duality1} \\
\lambda'   &=&(a\lambda+b)(c\lambda+d)^{-1},\label{duality22}  \\ 
\chi'      &=&(c\lambda+d)\chi (cS^{\dagger}+d)^{\dagger}.\label{duality3}
\eeqarr
To show the invariance of ${\rm Tr}[ i(\lambda-S)\chi]$
it is convenient to rewrite~\rref{duality1} as
\[
 S'=(cS^{\dagger}+d)^{-{\dagger}}(aS^{\dagger}+b)^{\dagger}.
\]
The proof of invariance of the remaining term which can be written as
\[
{\rm Re}\,{\rm Tr}\,[-\frac{i}{2} \lambda \chi S_2
  \chi^{\dagger}]
={\rm Tr}\,[\frac{1}{2}\lambda_2 \chi S_2 \chi^{\dagger}]~,
\]
is straightforward using the following transformations 
obtained from~\rref{duality1}, \rref{duality22} and \rref{duality3}
\beqarr
S_2'         &=&(cS^{\dagger}+d)^{-\dagger} S_2(cS^{\dagger}+d)^{-1}, 
\nonumber \\
\lambda_2'   &=&(c\lambda+d)^{-\dagger}\lambda_2(c\lambda+d)^{-1},
\label{duality2}  \\
\chi'^{\dagger}      &=&(cS^{\dagger}+d)\chi^{\dagger} 
(c\lambda+d)^{\dagger}~.
 \nonumber
\eeqarr

The Lagrangian  has also a discrete parity symmetry which acts on the
fields as
\beqarr
\alpha' &=& \bar{\alpha}~, \nonumber  \\
\beta' &=& -\bar{\beta}~, \nonumber   \\
S'        &=&   - \bar{S}~,  \label{parity}  \\
\chi'     &=&  \bar{\chi}~, \nonumber \\
\lambda'  &=&  - \bar{\lambda}~.  \nonumber 
\eeqarr 

Although the theory of duality invariance presented in the previous
section guarantees that this theory is self-dual, one can also check 
directly that the equations of motion 
obtained by varying the auxiliary fields are preserved
under duality rotations. These equations of motion are
\beqarr
\Lag_{\la}\equiv{{\partial\Lag}\over {\partial \la^{T}}} 
&=& 
i(\chi-\frac{1}{2}\chi S_2 \chi^{\dagger} +\alpha-i\beta)=0~,
\label{EQ}
\\
\Lag_{\chi}\equiv
{{\partial\Lag}\over {\partial \chi^{T}}} &=& 
i(\lambda - S - iS_2 \chi^{\dagger} 
\lambda_2) = 0~,\label{L/chi}
\eeqarr
and indeed 
these two equations form a multiplet under duality transformations.
Using the explicit forms~\rref{EQ} and~\rref{L/chi} one can check that 
\beqarr
\delta \Lag_{\la} &=& 
(C\lambda + D)\Lag_{\la} +  \Lag_{\la}
 (\lambda C + D^{\dagger})
+ \chi\Lag_{\chi}  C~, \nonumber \\
\delta \Lag_{\chi} &=& 
- (SC + D^{\dagger})\Lag_{\chi}-
\Lag_{\chi} (C\lambda +
D)~.
\nonumber
\eeqarr
Alternatively, one can obtain these equations directly from~\rref{CovScalars}.

\section{Elimination of the Auxiliary Fields}
\label{NoAux}

In this section we study the equation of motion~\rref{EQ} and 
attempt to solve for $\chi$. We conjecture the form the Lagrangian 
assumes after the elimination of the 
auxiliary fields. This form is a generalization of the well-known Born-Infeld
Lagrangian to more than one gauge field.
 
Using the equation of motion (\ref{EQ}) in
the Lagrangian (\ref{BIL}) we obtain
\beqn
\Lag={\rm Re}\,{\rm Tr}\,\left[-i S\chi\right]
={\rm Tr}\,[\,S_2\chi_1+S_1\chi_2\,]~,
\label{BI2}
\eeqn
where $\chi$ is now a function of $\alpha$, $\beta$ and $S_2$ that 
solves (\ref{EQ}).  
For $n=1$ we have to solve a second order
algebraic equation and we obtain 
\[
\chi =\frac{1-\sqrt{1+2S_2\alpha - S_2^2 \beta^2}}{S_2}+i\beta~.
\] 
Apart from the fact that the gauge fields are complex  
the result is the Born-Infeld Lagrangian
\beqn
\Lag=1-\sqrt{1+2 S_2\alpha - S_2^2 \beta^2} +S_1\beta~.
\label{LBI}
\eeqn
In fact, for $n=1$ we could have taken the gauge fields to 
be real even in the formulation with auxiliary fields as in~\cite{RT}, 
in which case the duality group becomes the $Sp(2,\IR)$ 
subgroup of $U(1,1)$ obtained by requiring $a$, $b$, $c$ and $d$ to 
satisfy~\rref{sp2n} and to be real.

We now study equation  (\ref{EQ})
for arbitrary $n$. First notice that (\ref{EQ}) can be simplified with
the following field redefinitions
\beqarr
\widehat{\chi}&=&R\chi R^{\dagger}~, \nonumber \\
\widehat{\alpha}&=&R\alpha R^{\dagger}~,  \label{hatted}\\
\widehat{\beta}&=&R\beta R^{\dagger}~, \nonumber
\eeqarr
where, as in~\rref{S2rr}, $R^{\dagger}R=S_2$.
The equation of motion for $\chi$ is then equivalent to
\beqn
\widehat{\chi}-\frac{1}{2}\widehat{\chi}\widehat{\chi}^{\dagger} 
+\alh-i\beh=0~.
\label{EQ'}
\eeqn
Breaking this equation into its hermitian and antihermitian parts we find
\beqarr
\widehat{\chi}_2 &=& \beh ~,\\
\widehat{\chi}_1 &=& \frac{1}{2}
(\widehat{\chi}_1^2-2\alh +\beh^2 +i[\beh,\widehat{\chi}_1])~.
\label{chi1}
\eeqarr
It is convenient to define
\[
X \;=\;1 - \widehat{\chi}_1~. \nonumber
\]
Then~\rref{chi1} is equivalent 
to the quadratic 
equation for the hermitian matrix~$X$
\beqn
X^2=1+2\widehat{\alpha}-\widehat{\beta}^2+i[\widehat{\beta},X]~.
\label{xEQ}
\eeqn
In terms of $X$ the Lagrangian~\rref{BIL} takes the form
\beqn
\Lag={\rm Tr}\,[\,1-X \; +S_1\be]~, \label{Lred}
\eeqn
where here $X$ is now a function of $\alh$ and $\beh$ that 
satisfies (\ref{xEQ}).

For $n=1$ the equation~\rref{xEQ} can be
solved trivially since it is a second order algebraic equation.
For arbitrary $n$, it becomes a matrix equations
whose closed form solution  does not seem to be known.
However we solved for $X$ as a 
power series in $\alh$ and $\beh$
\beqn
X=\sum_{m \geq 0} \frac{1}{m!} X^{m}~.
\label{Xexp}
\eeqn
Here $m$ refers to the combined power of  $\alh$ and $\beh$ in each term.
Then $X$ can be solved  perturbatively using the  recursion 
relation obtained from~\rref{xEQ} 
\[
X^0=1~,~~~X^1=\alh~,~~~X^2=-\widehat{\al}^2-\widehat{\be}^2+i[\beh,\alh]~,
\]
\beqn
\forall~ m>2~,~~~~2 X^m =
-\sum_{j=1}^{m-1} \left(\!\!\matc m\\j\emat\!\! \right) 
X^j X^{m-j}
+i\, m\, [{\beh},X^{m-1}]~.\label{Xrec}
\eeqn
The initial condition  for the recursion relation $X^0=1$
guarantees that the Lagrangian has a physical weak field limit.
We have not been able to solve explicitly the recursion 
relation~\rref{Xrec} and obtain $X$ to all orders. However, to obtain 
the action only ${\rm Tr}\,[X]$\, is needed. 

It was conjectured in~\cite{BMZ} and checked up to the sixth order
that inserting the solution of~\rref{xEQ} into~\rref{BIL} gives 
the following Lagrangian
\beqn
\Lag={\rm Tr}\; [\, 1 - 
\Salhbeh
\sqrt{1+2 \alh - \beh^2} +
S_1\beta\, ] ~.
\label{BI}
\eeqn 
The square root is to be understood in terms of its power series
expansion. The symmetrizer $\Salhbeh$ acts by symmetrizing each
monomial with respect to the $\alh$ and $\beh$ variables\footnote{To
avoid confusion, we remark here 
that the nonabelian 
Born-Infeld Lagrangian introduced in~\cite{AN,Tln} also involves a
symmetrized trace. However, while in~\cite{AN,Tln} the
symmetrization is in the nonabelian field strength here the
symmetrization is in
$\widehat{\alpha}$ and $\widehat{\beta}$.},  
and is normalized so that $\Salhbeh\,\sma{$\circ\, $} 
\Salhbeh=\Salhbeh~.$ 
It is a linear operator which maps a monomial of order $m$ in 
$\widehat{\alpha}$
and $\widehat{\beta}$ into $1/m!$\, 
times the polynomial obtained by summing all
$m!$ permutations of the monomial. 
Let $P_{r,s}(\alh,\beh)$\, be the symmetric polynomial of
order $r$ in $\widehat{\alpha}$\, and of order $s$ in
$\widehat{\beta}$. It is the sum of all the
$\left({}^{r+s}_{~\,r}\right)$\, different words of length $r+s$\, 
for which $r$ of the letters are $\widehat{\alpha}$\, and $s$ of the
letters are $\widehat{\beta}$\,.
We can write the following explicit formula for  $P_{r,s}(\alh,\beh)$\,
\beqn
P_{r,s}(\alh,\beh) = 
\frac{1}{r!s!}\left( \frac{\partial}{\partial \mu}\right)^r
\left( \frac{\partial}{\partial \nu}\right)^s (\mu \alh + \nu \beh)^{r+s}~.
\label{pdel}
\eeqn
Here $\nu$ and $\mu$ are commuting variables. 
Let $M^{rs}$ be an
arbitrary monomial with unit coefficient of order $r$ in
$\widehat{\alpha}$\, 
and of 
order $s$ in $\widehat{\beta}$\,. 
Then the symmetrizer acts on $M^{rs}$ as follows
\beqn
\Salhbeh(M^{rs})=
\left(\!\!\matc r+s\\r\emat \!\!\right)^{-1}\;P_{r,s}(\alh,\beh) ~.\label{PS}
\eeqn
The explicit form of the
symmetrized square root term appearing in the Lagrangian  is given by 
\[
{\Salhbeh}\,\sqrt{1+2{\alh}-{\beh}^2}~=~
\sum_{r,s\geq 0}(-1)^{(r+1)}\,~{(2r+2s-3)!!~(2s)!\over 2^{s}~(r+2s)!~s!}~
P_{r,2s}(\alh,\beh)~,
\]
where $(-3)!!=-1\,,~(-1)!!=1\,$. 

The conjecture~\rref{BI} can be sharpened by stating  
that the solution $X=X(\alh,\beh)$ of equation~\rref{xEQ}
satisfies
\beqn
{\cal C}\,(X)= {\Salhbeh}\,\sqrt{1+2{\alh}-{\beh}^2~}~,
\label{noproof}
\eeqn
where ${\cal C}$ is the cyclic average operator 
defined as follows:  ${\cal C}$ is a linear operator
and it maps a monomial 
of order $m$ in 
${\alh}$ and ${\beh}$ to $1/m$ times the polynomial 
obtained by 
summing all $m$ cyclic permutations of the 
monomial. 
The normalization of  ${\cal C}$ guarantees that 
${\cal C}\sma{$\,\circ\;$} {\cal C}={\cal C}$
and also 
Tr$\sma{$\,\circ\;$} {\cal C}=$ Tr\,.
Notice that the nontrivial statement in~\rref{noproof} is 
${\cal C}(X)=\Salhbeh\,(X)$\,, that is the cyclic average of $X$ is 
a completely symmetrized quantity. 
Using the sharper conjecture~\rref{noproof} 
it is straighforward to see that the 
Lagrangian~\rref{Lred}
\[
\Lag={\rm Tr}\,[\,1-X +S_1\be\;]={\rm Tr}\,[\,1-{\cal C}(X)+S_1\be \;]
\]
takes the form~\rref{BI}. We have no general analytic proof of~\rref{noproof}
but we have checked it up to order seventeen for arbitrary noncommuting
variables $\widehat{\alpha}$ and $\widehat{\beta}$ using the Mathematica
computer program. In the Appendix we present an alternative expansion, 
which is equivalent to~\rref{Xrec}, and which is more convenient for 
checking the conjecture~\rref{noproof} order by order by hand and by computer.

\section{Real field Strengths }
\label{RealF}

We now show that our results imply the existence of a Born-Infeld 
theory with $n$ real field strengths which is duality invariant under the
maximal duality group $Sp(2n,\IR)$. 

We first study the 
case without scalar fields, i.e.
$S_1=0$ and $S_2=R=1$.
Consider a Lagrangian $\Lc=\Lc(\al,\be)\,$ which describes a
self-dual theory  
with complex gauge fields. We will assume that the Lagrangian is a sum of
traces (or of products of traces) of monomials in $\alpha$\, and $\beta$\,.
It follows that this Lagrangian satisfies the self-duality 
equations~\rref{bselfdual}. This remains true in the special case that $\al$ 
and $\be$ are real. That is $\Lc=\Lc(\al,\be)$
satisfies the self-duality equation (\ref{bselfdual}) 
with $\al=\al^T=\bar{\al}$
and $\be=\be^T=\bar{\be}$. 
We now recall that equation (\ref{bselfdual}) is also 
the self-duality condition
for Lagrangians with real gauge fields provided that $\al$ and $\be$
are defined in the following way
\beqn
\alpha^{ab} = \frac{1}{4} F_{\!R}^{a} F_{\!R}^{b}~,
~~\beta^{ab} = 
\frac{1}{4}  \widetilde{F}_{\!R}^{a} F_{\!R}^{b}~,\label{realab}
\eeqn
where  $F^a_{\!R}$ denotes a real field strength.
This implies that the theory described by the Lagrangian 
$\Lc_{\!R}=\Lc(\al(F_{\!R}^a),\be(F_{\!R}^a))$ 
is self-dual with duality group $U(n)$, the 
maximal compact subgroup of  
$Sp(2n,\IR)$.
The duality group can be extended to the full noncompact $Sp(2n,\IR)$, 
the maximal duality group of $n$ real field strengths~\cite{GZ}, by
introducing the scalar fields $S$ via the prescription
(\ref{extension}) which also applies to the real case provided
$S$ is symmetric \cite{araki}.

In our case the Lagrangian
$
\Lag=\,{\rm Tr}\,[ 1-X(\widehat{\al},\widehat{\be})+S_1\be]~,
$
where $X(\widehat{\alpha},\widehat{\beta})$ 
is the solution of (\ref{xEQ}), 
defines a duality invariant theory because it is obtained from the 
Lagrangian with auxiliary fields (\ref{BIL}) that is explicitly 
self-dual.
Therefore
$\Lag_{\!R}={\rm Tr}\,[ 1-X(\widehat{\al},\widehat{\be}) + S_1\be]$ 
with the field strengths taken real
is also self-dual.
Using the conjecture (\ref{noproof}) we obtain an explicit formula for 
the Born-Infeld Lagrangian with real gauge fields describing 
an  $Sp(2n,\IR)$ duality invariant theory
\[
\Lag_{\!R}={\rm Tr}\; [\, 1 - 
{\cal{S}}_{_{\!\widehat{\al},\widehat{\be}}}\sqrt{1+2 \widehat{\al} - 
\widehat{\be}^2} +
S_1\beta\, ]~. ~
\]

\section{Supersymmetric Theory }
\label{SUSY}

In this section we briefly discuss supersymmetric versions of some of
the Lagrangians introduced. First we discuss the supersymmetric form
of the Lagrangian~\rref{BIL}.
Consider the superfields $V^a=\frac{1}{\sqrt{2}}( V^a_1+iV^a_2)$ 
and  $\check{V}^a=\frac{1}{\sqrt{2}}(V^a_1-iV^a_2)$ 
where $V^a_1$ and $V^a_2$ are real vector superfields, and define
\[
W^a_{\alpha}=-\frac{1}{4}\bar{D}^2 D_{\alpha} V^a~,~~
 \check{W}^a_{\alpha}=-\frac{1}{4}\bar{D}^2 D_{\alpha} \check{V}^a~.
\]
Both $W^a$ and $\check{W}^a$ are chiral superfields
and can be used to construct a matrix of chiral superfields
\[
{\cal M}^{ab}\equiv W^a \check{W}^b~.
\]
The supersymmetric version of the Lagrangian~\rref{BIL} is then given by
\beqn
\Lag={\rm Re}
\int \, d^2 \theta
\left[{\rm Tr}\,(
i(\lambda-S)\chi -\frac{i}{2} \lambda \bar{D}^2( \chi S_2 \chi^{\dagger})
- i  \lambda{\cal M}
)\right]~,\nonumber
\eeqn
where $S$, $\lambda$ and $\chi$ denote chiral superfields with the same
symmetry properties as their corresponding bosonic fields.
While the bosonic fields $S$ and $\lambda$ appearing in~\rref{BIL} are 
the lowest component of the superfields denoted by the same letter, 
the field $\chi$  in the action~\rref{BIL} is the
highest component of the superfield $\chi$.
A supersymmetric kinetic term for the scalar field $S$ can be written
using the K${\rm  \ddot{a}}$hler potential~\rref{kh} as described 
in~\cite{Zumino1}. 

Just as in the bosonic Born-Infeld, one would like to eliminate the
auxiliary fields. However we have not been able to do this exactly
except for $n=1$, and unlike the bosonic case we do not even have a
conjectured form of the Lagrangian without auxiliary fields.
For $n=1$ just as in the bosonic case the theory with auxiliary fields 
also admits both a real and a complex version, i.e. we can also consider a
Lagrangian with a single real superfield. Then we can
integrate out the auxiliary superfields and obtain the supersymmetric
version of~\rref{BI}  
\beqn
\Lag=\int \, d^4 \theta\frac{S_2^2 W^2\bar{W}^2}{1-A+\sqrt{1-2A+B^2}}
+
{\rm Re}\left[
\int \, d^2 \theta (-\frac{i}{2} S W^2)\right]~,
\label{sbi}
\eeqn
where
\[
A=\frac{1}{4}(D^2(S_2W^2)+\bar{D}^2(S_2\bar{W}^2))~,~~
B=\frac{1}{4}(D^2(S_2W^2)-\bar{D}^2(S_2\bar{W}^2))~.
\]
If we only want a $U(1)$ duality invariance we can set $S=i$ 
and then the action~\rref{sbi} reduces to the supersymmetric Born-Infeld action
described in~\cite{DP,CF,BG}. 

In the case of weak fields the
first term of~\rref{sbi} can be neglected and the Lagrangian is
quadratic in the field strengths. Under these conditions 
the combined requirements of
supersymmetry and self duality can be used~\cite{BG2} to constrain
the form of the weak coupling limit of the effective Lagrangian from
string theory.

\section{Extension to Arbitrary  Even Dimensions}
\label{forms}

In a space-time of arbitrary even dimension, $D=2p$ 
we define the matrices
\beqn
\alpha^{ab}= \frac{1}{p!}\,
F^{a}_{\mu_1\ldots \mu_p}\bar{F}^{b\,\mu_1\ldots \mu_p}~,~~
\beta^{ab}=\frac{1}{p!}\,
\tilde{F}^{a}_{\mu_1\ldots \mu_p}\bar{F}^{b\,\mu_1\ldots \mu_p}~,
\label{ab}
\eeqn 
where 
$\tilde{F}^{a}_{\mu_1\ldots \mu_p}=
1/p!~\varepsilon_{\mu_1\ldots \mu_{p}\nu_1\ldots \nu_{p}}
F^{a\, \nu_{1}\ldots \nu_{p}}$ is the Hodge dual of $F^a$.
The dual field strength is given by 
\[
\tilde{G}^a_{\mu_1\ldots \mu_p} 
\equiv p!\, \frac{\partial L}{\partial \bar{F}^{a\,\mu_1\ldots \mu_p}}~,~~
\tilde{\bar{G}}^a_{\mu_1\ldots \mu_p} 
\equiv p!\, \frac{\partial L}{\partial F^{a\,\mu_1\ldots \mu_p}}~.
\]
Since $\widetilde{\widetilde{F}}=(-1)^{p+1}F$ and 
$\tilde{F}G=(-1)^{p} F\tilde{G}$, 
for all even dimensions the matrix $\alpha$ is hermitian, while
$\beta$
is hermitian if $D=4\nu$ and anti-hermitian if  $D=4\nu+2$. It is also
convenient to define
\[
{\cal N}=\left\{\begin{array}{ll}
\alpha-i\beta ~,& {\rm if}~~D=4\nu~,   \\
\alpha+\beta ~,& {\rm if}~~D=4\nu+2~.
\emat
\right.  
\]
With these definitions the Lagrangian~\rref{BIL} gives a $U(n,n)$ duality
invariant theory in arbitrary even dimensions.

However, if the dimension of space-time is $D=4\nu+2$, where $\nu$ is
integer it is convenient 
to make the following field redefinitions
\[
\Lambda = i\lambda~,~~
{\cal S}= i S~.
\]
The new fields have the decomposition
\[
\Lambda=-\Lambda_1+\Lambda_2~,~~{\cal S}=-{\cal S}_1+{\cal S}_2~,
\]
where $\Lambda_1$ and ${\cal S}_1$ are hermitian and  
$\Lambda_2$ and ${\cal S}_2$ are anti-hermitian. The minus sign 
was introduced so that we have 
\beqn
{\cal S}_1=S_2~.
\label{S1S2}
\eeqn 
Then ${\cal S}_1$
is positive definite and  we
can write ${\cal S}_1 =R^{\dagger} R$\, with $R$ an arbitrary
nonsingular $n$-dimensional matrix. 

We also 
perform a similarity transformation on the $U(n,n)$ duality
group, such that the
transformation properties of the new fields simplify.
Let us define two $2n$-dimensional matrices with the block form
\[
\IK=\left(\begin{array}{rr}
0 & 1\\
-1 & 0
\emat
\right)~,~~
\IH=\left(\begin{array}{rr}
0 & 1\\
1 & 0
\emat
\right)~,
\]
and let the matrices $T$ and $M$ have the block decomposition
\[
M=\left(\begin{array}{rr}
a & b\\
c & d
\emat
\right)~,~~
T=\left(\begin{array}{rr}
\ul{a} & \ul{b}\\
\ul{c} & \ul{d}
\emat
\right)~.
\]
Then one can define the $U(n,n)$ group as the group of matrices
satisfying either one of the two relations 
\beqn
M \,\IK\, M^{\dagger} = \IK ~,~~
T\, \IH\, T^{\dagger} = \IH ~.
\label{unnrel}
\eeqn
The two definitions are related by a unitary transformation
 $M={\cal U}^{-1}T{\cal U}$ where
\beqn
{\cal U}=
\left(
\matcc
e^{i\pi/4} & 0 \\
0   & e^{-i\pi/4}
\emat
\right).
\label{Unitary}
\eeqn
The $n$-dimensional matrices $\ul{a}$, $\ul{b}$, $\ul{c}$ and
$\ul{d}$
satisfy
\beqn \ul{a}^{\dagger} \ul{d}+\ul{c}^{\dagger} \ul{b} =1 ~,~~
\ul{c}^{\dagger} \ul{a}+ \ul{a}^{\dagger} \ul{c}=0 ~,~~ \ul{b}^{\dagger} 
\ul{d} + \ul{d}^{\dagger} \ul{b}=0 ~.
\label{unn}
\eeqn

The action of $U(n,n)$ on the scalar fields is given by 
\beqarr
{\cal S}'         &=&(\ul{a}{\cal S}+\ul{b})
(\ul{c}{\cal S}+\ul{d})^{-1} ~, \nonumber \\
\Lambda'   &=&(\ul{a}\Lambda+\ul{b})(\ul{c}\Lambda+\ul{d})^{-1} ~,
\label{newduality}  \\
\chi'      &=&(\ul{c}\Lambda+\ul{d})\chi 
(-\ul{c}{\cal S}^{\dagger}+\ul{d})^{\dagger} ~.\nonumber
\eeqarr
Note that the positivity of ${\cal S}_1$ is 
compatible with the above transformation law of ${\cal S}$.

The Lagrangian, written in terms of the redefined fields, takes the form
\beqn
\Lag= {\rm Re}\left[{\rm Tr}\,(
(\Lambda-{\cal S})\chi -\frac{1}{2} \Lambda \chi {\cal S}_1 \chi^{\dagger}
+\Lambda {\cal N}
)\right].  \label{sp2nLag}
\eeqn
Our conjecture regarding the Lagrangian without auxiliary fields is
independent of the dimension of space-time and if it holds 
we can eliminate the auxiliary fields to obtain the Lagrangian 
\beqn
\Lag={\rm Tr\,[\,1-{\cal S_{\widehat{\alpha},\widehat{\beta}}}}\,
\sqrt{1+2\widehat{\alpha}+\widehat{\beta}^2}+
{{\cal S}_2\beta}\,]~, \label{yyy}
\eeqn
where 
\beqarr
\widehat{\alpha}&=&R \alpha R^{\dagger} ~,  \\
\widehat{\beta}&=&R \beta R^{\dagger} ~. \nonumber 
\eeqarr
Note also that ${\cal S}_2$ appears in the
last term of the Lagrangian~\rref{yyy}, and this is consistent with 
 ${\cal S}_2$ and $\beta$ being anti-hermitian
 in  space-times of odd half dimension. Also, 
there is a change of sign in front of the $\hat{\beta}^2$\, term
under the square root in~\rref{yyy} due to the change in the 
definition of ${\cal N}$. 

If the half-dimension of space-time is odd it is consistent to take
all the fields to be real in either the
Lagrangian with auxiliary fields~\rref{sp2nLag}, 
or in the Lagrangian~\rref{yyy} where the auxiliary fields have been 
eliminated. Then we obtain a theory invariant under 
an $O(n,n)$ duality group. 
It was shown in~\cite{tanii,araki} that  the
maximal connected duality group for a theory of dimension $D=4\nu +2$ with $n$ 
antisymmetric tensors is $SO(n,n)$. In the analysis of~\cite{tanii,araki} only
infinitesimal duality transformations were considered, and from these
one can only show duality under the connected component of the
group. 
In~\cite{Julia,Julia2} the group $O(n,n)$ was considered.
Note that, 
as discussed in Appendix B, $O(n,n)$ has four disjoint components 
embedded in $U(n,n)$ which is a connected group. 
Finally, one can also obtain a theory invariant
under the $O(n)\times O(n)$ maximal compact subgroup of $O(n,n)$ 
by setting ${\cal S}=- 1$ in the Lagrangian~\rref{yyy}.

\section*{Acknowledgments}

We would like to thank M. K. Gaillard for many illuminating
discussions. We are also grateful to Alexander Friedland and Jeffrey Anderson
for their help with the Mathematica computer program.
This work was supported in part by the Director, Office of Science,
Office of High Energy and Nuclear Physics, Division of High Energy
Physics of the U.S. Department of Energy under Contract
DE-AC03-76SF00098 and in part by the National Science Foundation 
under grant PHY-95-14797.
P.A. is supported by an INFN grant (concorso No. 6077/96). 

\section*{Appendix A}

In this appendix we discuss an equivalent expansion of the 
Lagrangian~\rref{BI2}
which simplifies  the order by order check of the conjecture. We
set $S=i$ for simplicity, since $S$ can always be reintroduced via the
prescription~\rref{extension}. 
The expansion is in terms of the variables $p$ and $q$ defined as
\beqn
p \equiv -\frac{1}{2}(\alpha - i\beta)~, ~~ 
q \equiv -\frac{1}{2}(\alpha +i\beta) ~.
\label{aandb}
\eeqn
Note that the self-duality
equation~\rref{bselfdual} simplifies when written in terms of $p$ and
$q$
\beqn
 p - \Lc_p\, p \,\Lc_p =  q- \Lc_q \, q \,\Lc_q~.
\label{selfdual}
\eeqn

Next we describe the perturbative expression of $X$.
Let us define
\[
\chi =  2P~ , ~~\chi^{\dagger} = 2Q~.
\]
Then the equations of motion~\rref{EQ} for $\chi$
and its hermitian conjugate 
become
\[
P = PQ + p~, ~~ Q = PQ + q~.
\]
It is convenient to consider the following expansions
\[
P = \sum_{n}P_n~,~~ Q = \sum_{n}Q_n~,
\]
we then  have
\[
P_0 = Q_0 = 0~, ~~P_1 = p~,~~Q_1 = q
\]
and we can solve for $P_n$ recursively 
\beqn
P_n = \sum_{r=1}^{n-1} P_r Q_{n-r}~.
\label{ABrec}
\eeqn
Notice that since ${\chi}-{\chi}^{\dagger}=2i\beta$~, 
for $n> 1$,  $P_n = Q_n~$.
Therefore we have $P_0=0~,~P_1=p~,~P_2=pq$\, and,  
for all $n>2$~,
\beqn
P_n=p P_{n-1}+P_{n-1}q+\sum_{r=2}^{n-2}P_r P_{n-r}~.\label{Arec}
\eeqn
We also have $P_n=-{1\over{2\,n!}} X^n$ for all $n>1$. 
The Lagrangian is now expressed as
\[
\Lc = {\rm Re}\,{\rm Tr}\,\chi =  {\rm Re}\,{\rm Tr}\,[\,2\sum_{n} P_n]
={\rm Tr}\, [\, p + q + 2\sum_{n \geq 2}P_n\,]~.
\]

Using~\rref{pdel} and the linear change of variables~\rref{aandb} one can
prove that symmetrization with respect to $\al$ and $\be$ is equivalent to 
symmetrization with respect to $p$ and $q$. Then 
we can rewrite
the conjectured symmetrized square root Lagrangian as
\beqn
\Lc ={\rm Tr}[1-{\Sab}\sqrt{\11 -2(p+q)+ (p-q)^2}\,]~.
\label{Lpq}
\eeqn

We believe the explicit power series 
expansion of the square root in~$p$ and~$q$ has
the simple expression
\[
{\Sab}\sqrt{\11 -2(p+q)+ (p-q)^2}=
\11-
p - q - 2 \sum_{r,s \geq 1} \frac{1}{r+s} 
\left( \!\!
\matc
r+s-2 \\
r-1
\emat  \!\!
\right)
P_{r, s}(p,q)~,
\]
which has been checked up to order twenty in
$p$ and $q$ with the Mathematica 
computer program.
Using the above expansion we can rewrite the conjecture~\rref{noproof} 
in the $p$ and $q$ variables
\[
{\cal C}\left( \,1-p - q - 2\sum_{n \geq 2}P_n\, \right)\,=1-
p - q - 2 \sum_{r,s \geq 1} \frac{1}{r+s} 
\left( \!\!
\matc
r+s-2\\
r-1
\emat \!\!
\right)
P_{r, s}(p,q)~.
\]
It is this form that has been checked up to order seventeen by computer.

\section*{Appendix B}

In this appendix we show that the field $S$ provides a global
parametrization of the coset space $G/H$ where $G$ is $U(n,n)$,
$Sp(2n,\IR)$ or $O(n,n)$ and $H$ is the maximal compact
subgroup of $G$. We will concentrate on $U(n,n)$ but the same
argument applies for the other   groups.

Cosets are equivalence classes of group elements $g$ of
$G$ under right multiplication with arbitrary elements $h$ of $H$
\[
g \sim gh~.
\]
We denote the coset containing $g$ by $gH$. The maximal compact subgroup
of $G$ is defined as
\[
H\equiv\{h \in G \,|\, hh^{\dagger} =h^{\dagger}h=1\}~.
\]
It is the intersection of $U(n,n)$ with $U(2n)$ i.e. $U(n)\times U(n)$.

Next consider the  map $\phi:G/H \rightarrow C$ 
defined by
\[
\phi(gH)=gg^{\dagger}~,
\]  
where 
\[
C=\{s\in G \,|\, s^{\dagger}=s,~s~{\rm positive~definite}
\}
\]
is the subset of
hermitian positive definite group elements of $G$.
This map is well defined since for any two elements $g$ and $g'$ in
the same coset, $g'=gh$ and $g'g'^{\dagger}
=ghh^{\dagger}g^{\dagger}=gg^{\dagger}$.
Furthermore this map is one to one. 
We show first that the map is surjective. Let
$s$ be an arbitrary  hermitian positive definite element of $G$. Then
\beqn
s=
\left(
\matcc
a & b \\
c & d
\emat
\right)
=
\left(
\matcc
1 & bd^{-1} \\
0 & 1
\emat
\right)
\left(
\matcc
d^{-\dagger} & 0 \\
0 & d
\emat
\right)
\left(
\matcc
1 & 0 \\
d^{-1}c & 1
\emat
\right)~.
\label{dec}
\eeqn
The last equality in~\rref{dec} can be checked using the group 
relations~\rref{sp2n}.
The decomposition exists whenever $d$ is
invertible, but since $s$ is positive definite and $d$ is
the restriction of  $s$ on an $n$-dimensional subspace $d$
is also positive definite. 
Note also that $d^{\dagger}=d$ and $(bd^{-1})^{\dagger}=bd^{-1}=d^{-1}c$.
Then $g$ defined as
\[
g=
\left(
\matcc
1 & bd^{-1} \\
0 & 1
\emat
\right)
\left(
\matcc
d^{-1/2} &  \\
0 & d^{1/2}
\emat
\right)
\]
satisfies $s=gg^{\dagger}$, thus the map $\phi$ is surjective.
To show that the map is also injective note that 
$gg^{\dagger}=g'g'^{\dagger}$ is equivalent to 
$g'^{-1}g(g'^{-1}g)^{\dagger}=1$. Then $h=g'^{-1}g$ is an element of
$G$ satisfying $hh^{\dagger}=1$, that is it belongs to the maximal
compact subgroup H and we have $g=g'h$ so  $g$ and $g'$ belong to the
same coset.

If we define $S_2=d^{-1}$ and $S_1=bd^{-1}$ we can rewrite~\rref{dec}
as
\[
s=
\left(
\matcc
1 & S_1 \\
0 & 1
\emat
\right)
\left(
\matcc
S_2 & 0 \\
0 & S_2^{-1}
\emat
\right)
\left(
\matcc
1 & 0 \\
S_1 & 1
\emat
\right)~.
\]
This decomposition can also be written in terms of $S_2$ and $S=S_1+i
S_2$ as
\beqn
s=
i
\left(
\begin{array}{rr}
0&1\\
-1&0
\emat
\right)
+
\left(
\matcc
0 & S^{\dagger} \\
0 & 1
\emat
\right)
\left(
\matcc
S_2^{-1} & 0 \\
0 & S_2^{-1}
\emat
\right)
\left(
\matcc
0 & 0 \\
S & 1
\emat
\right)~.
\label{sdec}
\eeqn
Left multiplication on the group $G$
induces an action of the group $G$ on the coset 
space
\[
s'= 
\left(
\matcc
a & b\\
c & d
\emat
\right)
s 
\left(
\matcc
a & b\\
c & d
\emat
\right)^{\dagger}.
\]
Using the decomposition~\rref{sdec} one can easily show that the
fractional transformation~\rref{dualSS} 
of $S$ is equivalent to this action. The form~\rref{sdec} is
very convenient  since the first term is invariant under the action,
while the second term only contains $S$ and $S_2$ and these have the simple
transformation properties~\rref{dualSS} and~\rref{S2trans}.

If we make all the matrices above real we obtain the parametrizations
of $Sp(2n,\IR)/U(n)$. If we  change the basis with the unitary 
matrix ${\cal U}$ defined in~\rref{Unitary} and then
require all the matrices to be real we 
obtain the coset space $O(n,n)/O(n)\times O(n)$.

Since the map $\phi$ is injective we see that $S$, such that $S_2$ is
positive definite, is a global coordinate on the coset space
$U(n,n)/U(n)\times U(n)$. Thus this coset space is connected. The group
$U(n,n)$ is a principal bundle over  $U(n,n)/U(n)\times U(n)$ with 
a $U(n)\times U(n)$ fiber. The number of disconnected components
of a principal bundle with a connected base is at most equal to the
number of components of the fiber which in this case is one. Thus
$U(n,n)$ is connected. Using the same argument one can show that
$Sp(2n,\IR)$ is connected while $O(n,n)$ has at most four
components. By  an argument similar to the one
used for the Lorentz  group one can show that there are at least four
components. Thus, as mentioned in Section~\ref{forms},  $O(n,n)$
has exactly four components.

\end{document}